\documentclass[%
 reprint,
superscriptaddress,
 amsmath,amssymb,
 aps,
prb,
longbibliography,
]{revtex4-1}

\usepackage{graphicx}
\usepackage{dcolumn}
\usepackage{bm}
\usepackage{hyperref}

\usepackage{braket}
\usepackage{xcolor}
\usepackage{soul}

\begin{document}

\preprint{APS/123-QED}

\title{SrCu$_2$(BO$_3$)$_2$ under pressure: a first-principle study}

\author{Danis I. Badrtdinov}
\affiliation{Theoretical Physics and Applied Mathematics Department, Ural Federal University, 620002 Yekaterinburg, Russia}
\affiliation{Institute of Physics, Ecole Polytechnique F\'ed\'erale de Lausanne (EPFL), CH 1015 Lausanne, Switzerland}

\author{Alexander A. Tsirlin}
\affiliation{Theoretical Physics and Applied Mathematics Department, Ural Federal University, 620002 Yekaterinburg, Russia}
\affiliation{Experimental Physics VI, Center for Electronic Correlations and Magnetism, Institute of Physics, University of Augsburg, 86135 Augsburg, Germany}

\author{Vladimir V. Mazurenko}
\email{vmazurenko2011@gmail.com}
\affiliation{Theoretical Physics and Applied Mathematics Department, Ural Federal University, 620002 Yekaterinburg, Russia}

\author{Fr\'ed\'eric Mila}
\email{frederic.mila@epfl.ch}
\affiliation{Institute of Physics, Ecole Polytechnique F\'ed\'erale de Lausanne (EPFL), CH 1015 Lausanne, Switzerland}

\date{\today}

\begin{abstract}
Using density-functional theory (DFT) band-structure calculations, we study the crystal structure, the lattice dynamics, and the magnetic interactions in the Shastry-Sutherland magnet SrCu$_2$(BO$_3$)$_2$ under pressure, concentrating on experimentally relevant pressures up to 4\,GPa. We first check that a ferromagnetic spin alignment shortens the nearest-neighbor Cu-Cu distance and reduces the Cu-O-Cu angle compared to the state with the antiferromagnetic spin alignment in the dimers, in qualitative agreement with the structural changes observed at ambient pressure as a function of temperature and applied field. Next, we determine the optimal crystal structures corresponding to the magnetic structures respectively consistent with the dimer phase realized at ambient pressure, the N\'eel ordered phase realized at high pressure, and two candidates for the intermediate phase with two types of dimers and different stackings. For each phase, we performed a systematic study as a function of pressure, and we determined the exchange interactions and the frequencies of several experimentally relevant phonon modes. In the dimer phase, the ratio of the inter- to intra-dimer couplings is found to increase with pressure, in qualitative agreement with various experiments. This increase is mostly due to the decrease of the intra-dimer coupling due to the reduction of the Cu-O-Cu angle under pressure. The phonon frequency of the pantograph mode is also found to increase with pressure, in qualitative agreement with recent Raman experiments. In the N\'eel phase, the frequency of the pantograph mode is larger than the extrapolated value from the dimer phase, again in agreement with the experimental results, and accordingly the intradimer coupling is smaller than the extrapolated value from the dimer phase. Finally, all tendencies inside the candidate intermediate phases are thoroughly worked out, including specific predictions for some Raman active phonon modes that could be used to pin down the nature of the intermediate phase.
\end{abstract}

\maketitle


\section{\label{sec:introduction}Introduction}

\begin{figure}[!h]
\includegraphics[width=0.4\textwidth]{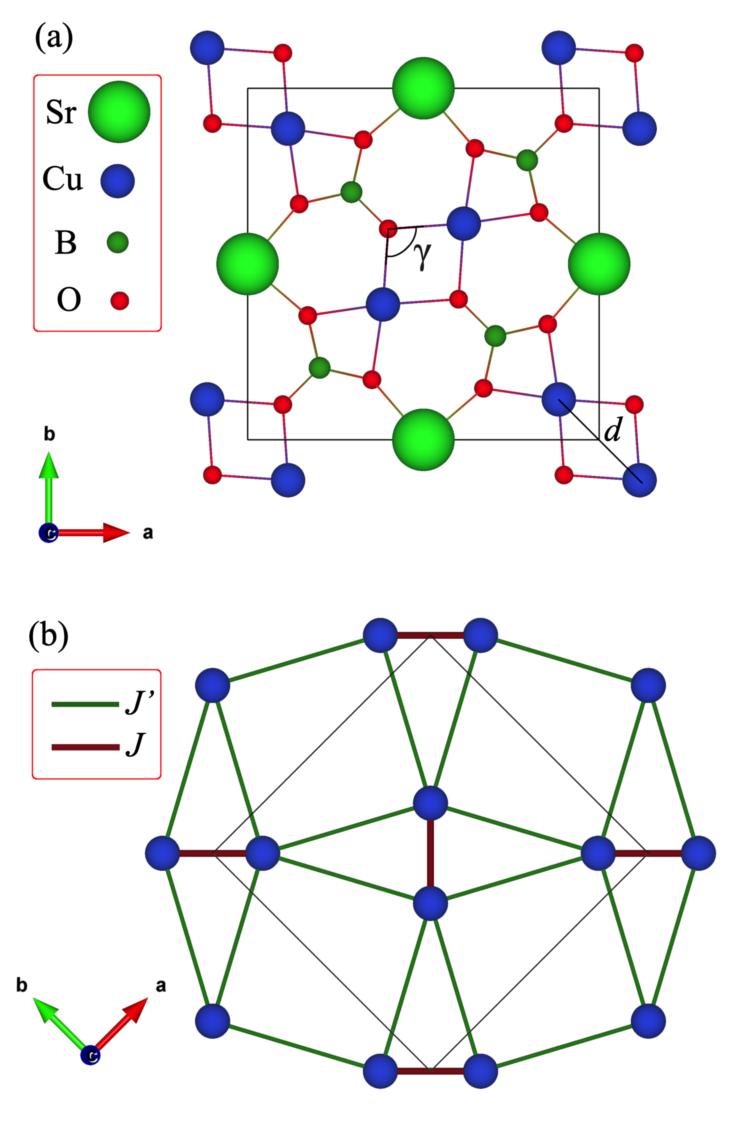}
\caption{a) Crystal structure of SrCu$_2$(BO$_3$)$_2$. $d$ is the shortest distance between two magnetic Cu$^{2+}$ ions, and $\gamma$  is the Cu -- O -- Cu angle  within a dimer.  (b) Shastry-Sutherland model mapped on the corresponding crystal structure with intradimer $J$ and interdimer $J^{\prime}$ exchange interactions.   The \texttt{VESTA} software was used for the crystal structure visualization~\cite{vesta}.}
\label{fig:Crystal}
\end{figure}

\begin{figure*}[]
\includegraphics[width=0.8\textwidth]{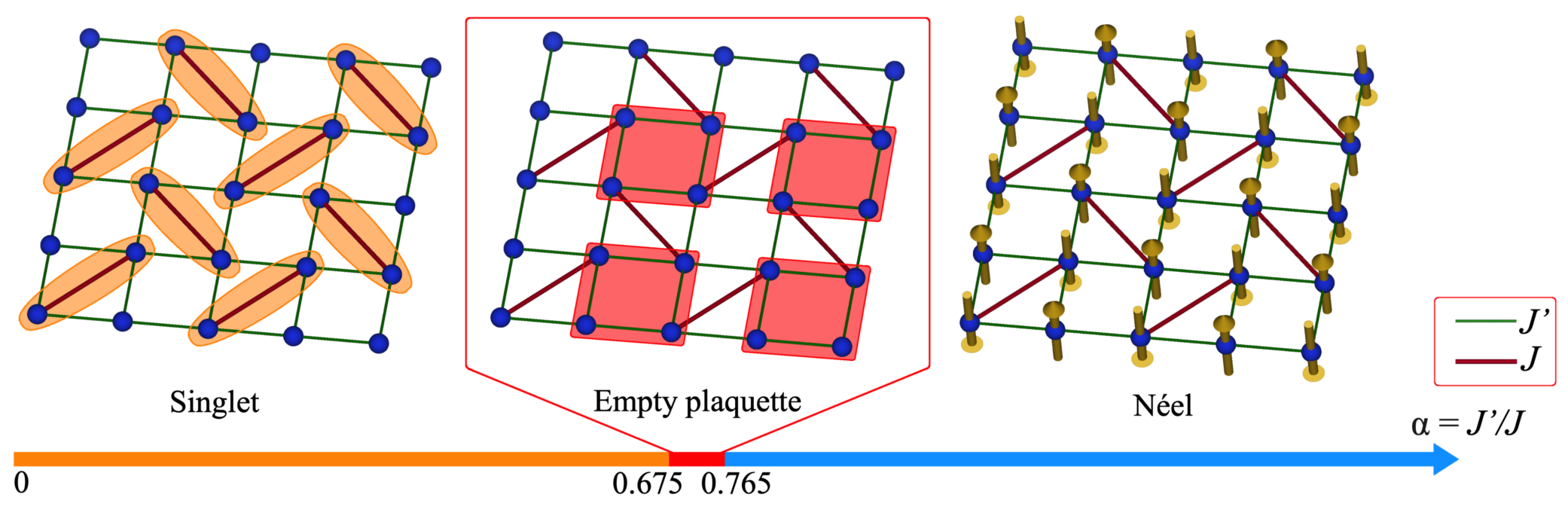}
\caption{Phase diagram of the Shastry-Sutherland model with intradimer $J$ and interdimer $J^\prime$ exchange interactions. Varying their ratio $\alpha = J^\prime/J$ leads to phase transitions at the critical values $\alpha_{c1} $ = 0.675 and  $\alpha_{c2} $ = 0.765  from dimer singlets (orange ovals) to plaquette state (red squares) and from plaquette to ordered N\'eel state, respectively~\cite{Corboz}.}
\label{fig:SS_model}
\end{figure*}

After being one of the main players in the field of frustration-induced magnetization plateaus, strontium copper borate SrCu$_2$(BO$_3$)$_2$ is now systematically studied under pressure. This compound consists of well separated  layers of  [CuBO$_3$]$^{-}$ structural elements stacked along the $c$ axis. Each magnetic Cu$^{2+}$ ion carries a spin $S$ = 1/2, leading to a 2D arrangement of mutually orthogonal dimers (Fig.~\ref{fig:Crystal}), a model topologically equivalent to the Shastry-Sutherland (SS) model~\cite{SS} (see Fig.~\ref{fig:SS_model}) defined by the Hamiltonian:    
\begin{eqnarray}
\hat{H} =  J \sum_{\braket{i,j} }\hat{ \mathbf{S}}_i \cdot \hat{ \mathbf{S}}_j  + J^{\prime} \sum_{\braket{ {\braket{i,j}} } }  \hat{ \mathbf{S}}_i \cdot \hat{ \mathbf{S}}_j 
\label{eq:Ham}
\end{eqnarray}
The Cu$^{2+}$ spins of a dimer are coupled by an intra-dimer exchange integral $J$ involving a Cu -- O -- Cu path, while the spins of the neighboring dimers interact through an inter-dimer  coupling $J^\prime$. In the limit  $\alpha = J^\prime / J \rightarrow 0 $, the ground state is an exact product of dimer singlets and the spectrum is gapped,  while in the $\alpha  \rightarrow  \infty$ limit, the model reduces to the standard Heisenberg model on the square lattice, with long-range antiferromagnetic order and a gapless spectrum. In between, after some debate, it is now well accepted, based on numerical evidence, that there in an intermediate plaquette phase in the range  $\alpha$ =  0.675 $\div$ 0.765~\cite{Corboz, Koga2000, Takushima2001, Sachdev2019, Nakano2018, Miyahara}. In this phase, strong bonds form around empty plaquettes, leading to a two-fold degenerate ground state with broken translational symmetry and a gapped spectrum. Various experiments have shown that, at ambient pressure, the parameter $\alpha$ is around 0.63 in SrCu$_2$(BO$_3$)$_2$, putting the system into the dimer phase, but not too far from the intermediate plaquette phase~\cite{Miyahara}. This has been confirmed by ab-initio calculations done for the ambient-pressure crystal structure~\cite{mazurenko2008, Pantograph}. 

To explore this phase diagram experimentally, it has been proposed early on to use pressure to increase the ratio $\alpha$, and indeed a phase transition around 2 GPa has been detected first by NMR~\cite{Waki2007, Takigawa2010} and then confirmed by inelastic neutron scattering (INS)~\cite{Zayed}, magnetization measurements~\cite{Haravifard2012}, electron spin resonance (ESR)~\cite{Sakurai}, thermodynamic measurements~\cite{Sandvik}, and Raman scattering~\cite{Zheludev2019}. However, the properties of this phase do not seem to be consistent with the plaquette phase predicted by the numerical results obtained on the Shastry-Sutherland model. In particular, NMR reveals the presence of two different Cu sites in an applied field, whereas in the plaquette phase of the Shastry-Sutherland model all magnetic sites remain equivalent. Even in zero field there seems to be a problem with the plaquette phase. Indeed, the structure factor revealed by neutron scattering seems to be more consistent with another plaquette  phase in which strong bonds form around plaquettes with diagonal couplings, a phase naturally leading to two types of Cu sites. 

On the theory side, it has been shown that this phase with two different copper sites can be stabilized by a distortion that would lead to two types of inter- and intra-dimer couplings, and to be essentially one-dimensional~\citep{Moliner, Boos}. Accordingly, it has been called the Haldane or full plaquette phase (FPP) (Fig.~\ref{fig:Haldane}), by contrast to the empty plaquette phase (EPP) of the Shastry-Sutherland model (Fig.~\ref{fig:SS_model}). In the following, we will refer to this phase as the FPP/Haldane phase.

To make progress on the properties of SrCu$_2$(BO$_3$)$_2$ under pressure, and in particular on the identification of the intermediate phase, it is of utmost importance to have more precise information on the structure of the system, and on the exchange couplings that describe its magnetic properties, as a function of pressure. The goal of the present paper is to address this issue with ab-initio calculations, the only ab-initio calculations available so far being, to the best of our knowledge, limited to ambient pressure~\cite{mazurenko2008, Pantograph}.

In the present paper, we study the structural properties and the corresponding exchange interactions of SrCu$_2$(BO$_3$)$_2$, their evolution under pressure and with different magnetic order by using density functional theory (DFT) techniques. The remainder of this paper is organized as follows.  Sec.~\ref{sec:methods} covers the methods used in our calculations. In Sec.~\ref{sec:Tetragonal}, we consider magnetic orders which do not break the initial tetragonal symmetry of the system, and we provide a comprehensive comparison of  the calculated structural parameters,  exchange couplings and phonon modes with available experimental data.   Moreover, we also demonstrate the  stabilization of the so called FPP/Haldane magnetic state~\cite{Moliner, Boos}, which can be considered as the main candidate for intermediate state, as discussed in Sec.~\ref{sec:Haldane}.  Finally, we summarize our findings and discuss possible ways to detect this magnetic phase in experiments in Sec.~\ref{sec:conclusions}.

\section{\label{sec:methods}Computational methods and models}
 {\it Electronic structure calculations.}
 Density-functional band-structure calculations for SrCu$_2$(BO$_3$)$_2$ were performed within the generalized gradient approximation (GGA) using the Perdew-Burke-Ernzerhof (PBE) exchange-correlation functional~\cite{pbe96}  as implemented in the Vienna $ab$ $initio$ simulation package \texttt{VASP}~\cite{vasp1, vasp2} with the plane-wave basis set. In these calculations, we set the energy cutoff to 920 eV and the energy convergence criteria to 10$^{-7}$ eV. For the Brillouin zone integration a 4$\times$4$\times$4 Monkhorst-Pack mesh was used. In order to check the possible convergence errors and structural relaxation error arising  due to the incompleteness of the basis set known as Pulay stress~\cite{Pulay}, all calculations were compared with those obtained using a 1100 eV energy cutoff and a 6$\times$6$\times$6  k-mesh. Correlation effects were taken into account within the static mean-field DFT+$U$ approach~\cite{anisimov1991} with fully localized limit  (FLL) for the double counting correction.  
 
{\it Structural optimization.}
The crystal structure of SrCu$_2$(BO$_3$)$_2$ at low temperature and ambient pressure was taken from Ref.~\onlinecite{Zayed_PhD_diss}. In accordance with experimental studies~\cite{Zayed, Sandvik, Zheludev2019} we consider pressures ranging from 0 to 4 GPa.  Both the lattice parameters and the atomic positions were allowed to relax under fixed hydrostatic pressure until all the residual force components of each atom were less than 5 $\times$ 10$^{-4}$ eV/\AA. The convergence in interatomic distances, exchange interactions, and phonon frequencies is 10$^{-4}$\AA, 1 K, and 0.1 cm$^{-1}$, respectively. All these parameters may deviate from the experimental values due to systematic errors in DFT, but their relative changes between the different magnetic configurations or as a function of pressure are significant and robust, as we show below.

The symmetry of the optimized structures was analyzed by the \texttt{FINDSYM} software~\cite{Findsym}.  We used the finite displacement method for the calculation of phonon spectra at the $\Gamma$  point and the \texttt{SAM} application on the Bilbao Crystallographic Server~\cite{Bilbao}  to analyse the symmetry of the calculated modes.

{\it Magnetic model.}
The exchange interaction parameters $J$ and $J^\prime$ of the spin Hamiltonian, Eq.~\ref{eq:Ham}, were extracted from the total energy of ordered collinear spin configurations using the mapping procedure~\cite{Xiang} within the DFT+$U$ method which, in the simplest case of two spins, can be written as
\begin{eqnarray}
J =(E_{FM} - E_{AFM})/{2S^2},
\end{eqnarray}
where $E_{FM}$ and $E_{AFM}$ denote the energy of the system in the ferromagnetic and antiferromagnetic configurations, respectively.  The spin-lattice coupling is taken into account implicitly by relaxing the crystal structures prior to calculating the exchange couplings. 

{\it Electronic model.} In order to give a microscopic explanation of the obtained inter-atomic exchange interactions, we perform a parametrization of the DFT band spectra of SrCu$_2$(BO$_3$)$_2$ around the Fermi level by constructing maximally localized Wannier functions, as implemented in the \texttt{wannier90} package~\cite{wannier90}. More specifically, we construct the following tight-binding model:
\begin{eqnarray}
\hat{H}_{TB} =  \sum_{i j,\sigma}t_{ij} \hat{c}^{+}_{i \sigma} \hat{c}_{j \sigma},
\end{eqnarray}
where $t_{ij}$ is the hopping integral between the i$th$ and j$th$ Wannier functions. On the basis of these Wannier functions, one can perform a microscopic analysis of the exchange interactions by using a superexchange theory approach~\cite{Cu2GeO4}: 
\begin{eqnarray}
J_{ij} =  J^{kin}_{ij} + 2J^{F}_{ij} = \frac{4 t^2_{ij}}{\widetilde{U}_{ij}}+ 2J^{F}_{ij}.
\label{eq:Coupling}
\end{eqnarray}
Here $\widetilde{U}_{ij} = U_{ii} - U_{ij}$ is an effective screened Coulomb parameter. The first term has a superexchange origin~\cite{anderson1959}, whereas the second term $J^{F}_{ij}$, the so-called non-local exchange interaction, arises due to the direct overlap of the wave functions~\cite{Rosner2002, Cu2GeO4}. All these parameters can be estimated by using the parametrization of the electronic structure obtained with first-principles calculations, which will be discussed below.

{\it Adjusting the DFT+U parameters.}
SrCu$_2$(BO$_3$)$_2$, as well as other Cu$^{2+}$ oxide materials~\cite{Prishchenko, danis2016, mazurenko2008} with partially occupied 3$d$ energy bands, belongs to the family of correlated insulators. Its ground-state properties, including magnetic moments and energy band gap, should be described by taking into account Coulomb correlations in the 3$d$ states of copper atoms. In our case, we use the DFT+$U$ method~\cite{anisimov1991}. Such an approach requires the choice of the on-site Coulomb $U$ value, which can be different for different transition-metal oxides. It is a common practice to consider $U$ as an empirical parameter, and to adjust it to better fit experimental data. Consequently, the results of the structural optimization will depend not only on the specific magnetic order under consideration, but also on the value of $U$. In this work, we fix $U$ first and then we determine the optimized crystal structures for different types of magnetic order.
 
\begin{figure}[!h]
\includegraphics[width=0.49\textwidth]{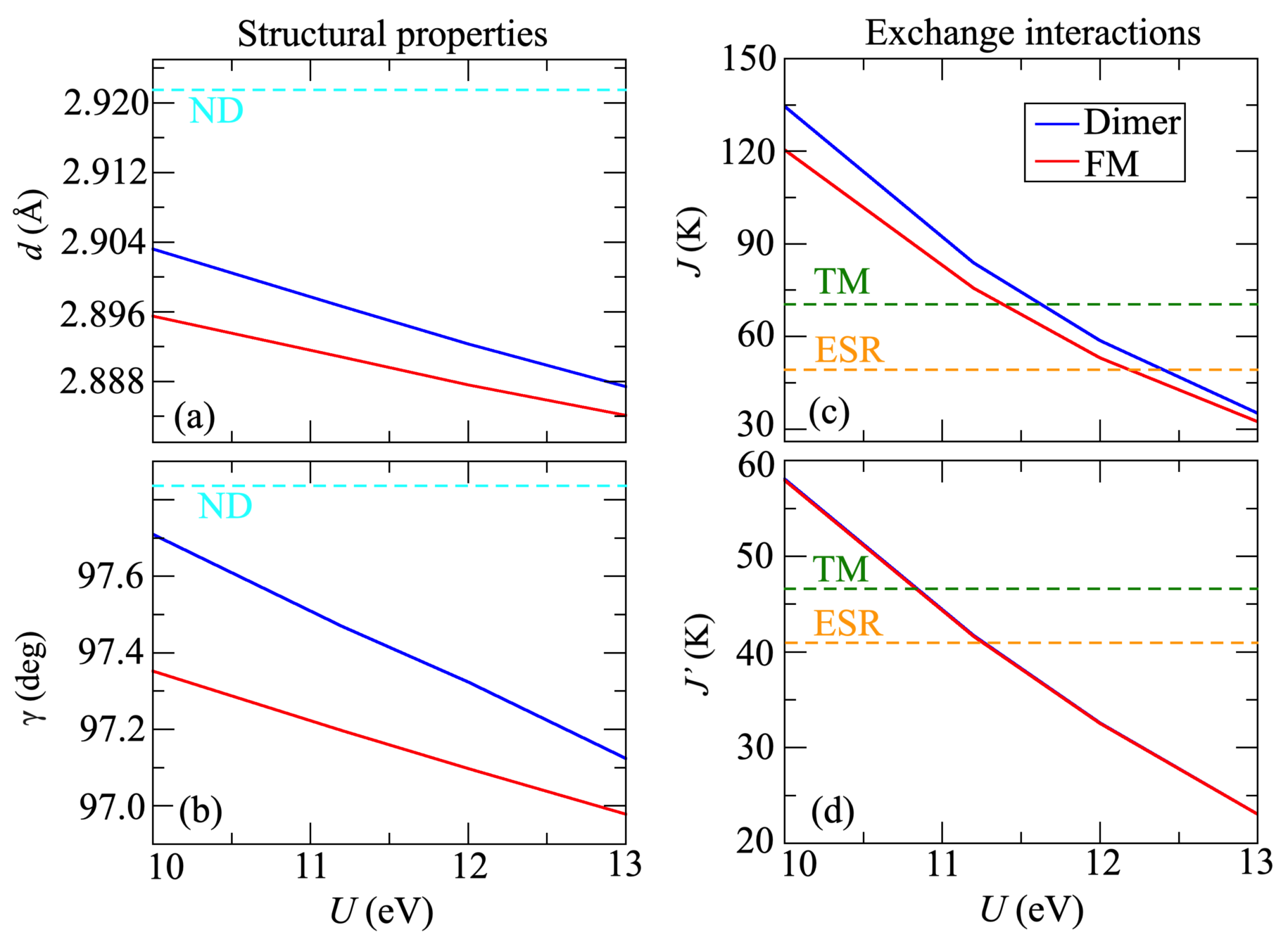}
\caption{Dependence of the ambient-pressure structural properties and magnetic couplings of SrCu$_2$(BO$_3$)$_2$ on the value of the on-site Coulomb $U$ parameter of DFT+$U$. Red and blue lines correspond to the calculations with ferromagnetic and Dimer orderings (Fig.~\ref{fig:Order}), respectively. (a) Intradimer distance $d$; (b) Cu -- O -- Cu $\gamma$ angle; (c) intradimer $J$ and  (d) interdimer  $J^\prime$ exchange interactions. The structural data from neutron diffraction (ND)~\citep{Zayed_PhD_diss}, experimental estimates of the exchange couplings from thermodynamic measurements (TM)~\cite{Zayed} and electron spin resonance (ESR)~\cite{Sakurai} experiment under ambient pressure are given for comparison. } 
\label{fig:U_variation}
\end{figure} 
 
In our study, we performed calculations with several values of $U$, as shown in Fig.~\ref{fig:U_variation}. As one could expect, increasing $U$ amplifies the localization of the Cu $3d$ electrons, as revealed by analyzing the magnetic moments of the copper and oxygen atoms. The choice of the $U$ value also affects the resulting exchange interactions and the corresponding structural properties. We emphasize that after comparing the exchange interactions calculated for different $U$ with available experimental estimates~\cite{Sakurai, Zayed}, we choose $U$ = 11.2 eV as the value that gives the best agreement between experimental and theoretical exchange couplings at ambient pressure. This value will be used to describe the magnetic behavior of SrCu$_2$(BO$_3$)$_2$ under pressure. Note that a similar value of the on-site Coulomb $U$ was previously used to study another copper oxide system~\cite{danis2016}, for which a reasonable agreement between calculated thermodynamics quantities, such as magnetization and magnetic susceptibility, and experimental data was reported.

The intra-atomic exchange interaction $J_H$, another important parameter of the DFT+$U$ scheme, was fixed at 1 eV. This value is chosen in a semi-empirical manner on the basis of previous works on Cu$^{2+}$ oxide materials ~\cite{Prishchenko, danis2016, mazurenko2008}, in which the DFT+$U$ results agree with available experimental data.

\section{\label{sec:Tetragonal} Tetragonal structures }

\begin{figure}[!h]
\includegraphics[width=0.49\textwidth]{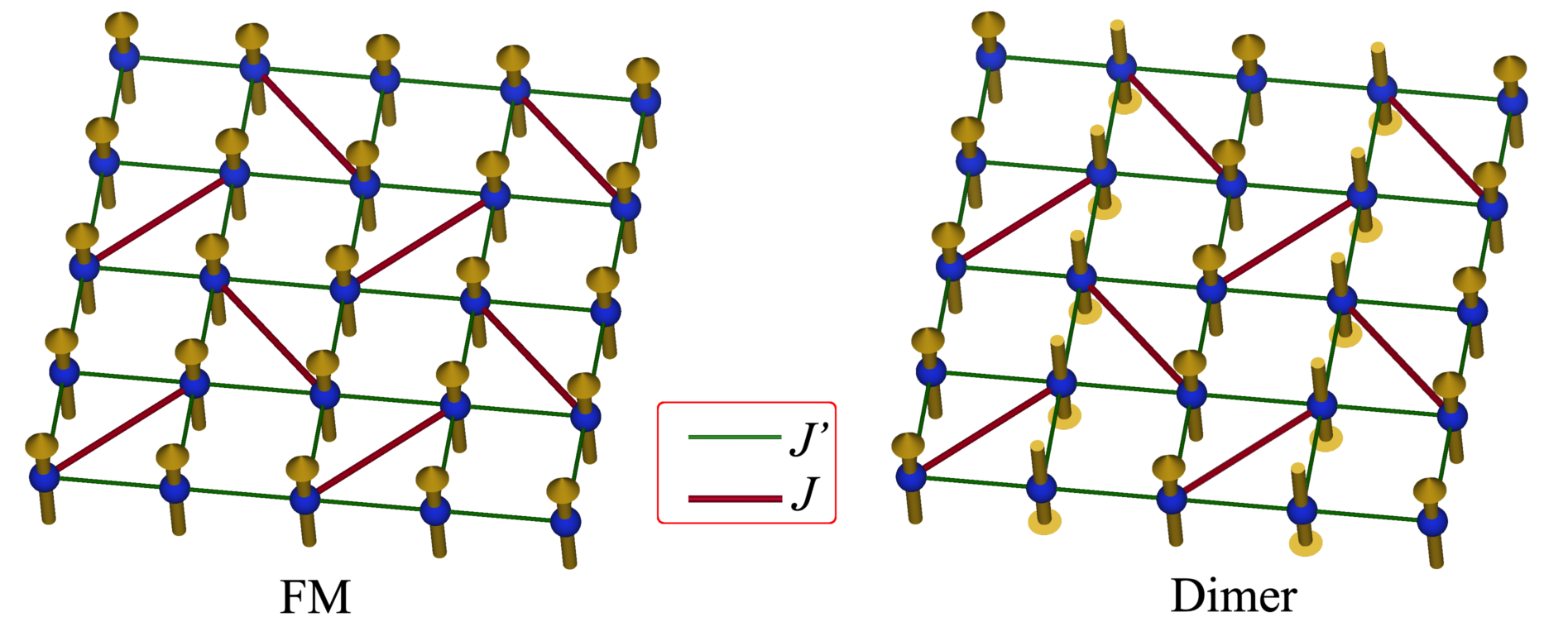}
\caption{Ferromagnetic and Dimer configurations of spins in the Shastry-Sutherland model used in this work to simulate the magnetic properties of SrCu$_2$(BO$_3$)$_2$.}
\label{fig:Order}
\end{figure}

The main focus of our study is on the magnetic properties of  SrCu$_2$(BO$_3$)$_2$, considering first a description in terms of the Shastry-Sutherland model. The ground state of SrCu$_2$(BO$_3$)$_2$ at ambient pressure is a product of dimer singlets. This is a many-body magnetic state that cannot be reproduced within a one-particle numerical approach such as the GGA approximation of the density functional theory we used. A standard practice in this situation  is to define a mean-field  magnetic state characterized by an antiferromagnetic configuration of the spins belonging to the dimer, in agreement with Ref.~\onlinecite{Pantograph}. We will further refer to such a configuration as Dimer. We also consider other types of magnetic configurations, for instance the ferromagnetic (FM) alignment of the copper spins (Fig.~\ref{fig:Order})) that can be thought as a mean-field representation of the many-body triplet state induced by an external magnetic field~\cite{Matsuda2013}.The interatomic distances and exchange couplings pertinent to this state are denoted by $d^{Dimer/FM}$ and $J^{Dimer/FM}$, respectively.  Additionally, we consider N\'eel magnetic order, which is the only long-range-ordered state realized in the quantum version of the Shastry-Sutherland model (Fig.~\ref{fig:SS_model}). All these three magnetic states do not break the initial tetragonal symmetry of SrCu$_2$(BO$_3$)$_2$ in the sense that all the intra-dimer couplings in the system remain the same. In Sec.~\ref{sec:Haldane}, we will consider the case where intra-dimer couplings become different due to specific magnetic orders, consequently causing structural distortions.

\subsection{Ambient pressure}
Let us discuss the structural and magnetic characteristics at ambient pressure first.  According to the experimental data~\cite{Vecchini}, at ambient pressure the intradimer  Cu -- Cu  distance $d$ and  the Cu -- O -- Cu angle $\gamma$ change in the temperature range where spin-spin correlations set in.   Below 40\,K, both $d$ and $\gamma$, as well as the lattice parameter $a$ increase ($\frac{\Delta a}{a} |_{\Delta T = 40 K}$ = $1.7  \times 10^{-4}$)~\cite{Vecchini}.  At the same time,  in high magnetic fields the lattice parameter $a$ becomes smaller ($\frac{\Delta a}{a} |_{\Delta H = 50 T}$ = $ - 1.4  \times 10^{-4}$)~\cite{High_H, Pantograph}, which can lead to a reduction in $d$ and $\gamma$.  It means that the formation of the triplet state shrinks the dimer, whereas the formation of the singlet state expands it. This effect can be explained by a significant spin-lattice coupling~\cite{Kodama2002}. Since we simulate the many-body singlet and triplet states with mean-field Dimer and ferromagnetic configurations at the DFT level, it is interesting to compare the optimized structural parameters and exchange couplings for these two distinct magnetic orders.  

At ambient pressure, the difference in the intradimer  Cu -- Cu  distance $d$ between the Dimer and FM configurations is $\Delta d = d^{Dimer } - d^{FM} \sim 0.006$ \AA ,\ while the difference in the Cu -- O -- Cu  $\gamma $ angle  $\Delta \gamma  \sim 0.28^\circ$, which are in reasonable agreement with previous ab-initio results obtained using another DFT code ($\Delta d$ = 0.009 \AA \, and $\Delta \gamma$ = 0.43$^\circ$)~\cite{Pantograph}.

The experimental estimates allow us to perform a similar analysis between the low-temperature singlet and high-temperature paramagnetic states~\cite{Vecchini}: $\Delta d =  d^{T = 2 K } - d^{T = 40 K} \sim 0.013$ \AA ,\ and  $ \Delta \gamma = \gamma^{T =2 K } -  \gamma^{T = 40 K} \sim 0.5^\circ$.  Comparing $\Delta d$ and $\Delta \gamma$ we conclude that the theoretical parameters are in a reasonable range.   

On the basis of the optimized unit cell, we have further calculated the corresponding values of the $J$ and $J^\prime$ exchange couplings.  As one can expect from the structural information, the Dimer order demonstrates a stronger intradimer  $J$ coupling, in agreement with the Goodenough-Kanamori rule:  $J^{FM}$ = 75.6 K,    $J^{Dimer}$ = 83.8 K,   the ratio $J^{FM} / J^{Dimer}$ = 0.90.  On the other hand, the interdimer $J^\prime$ couplings are almost the same for each magnetic configuration, $J^{\prime FM}  = J^{\prime Dimer}$ = 41.7 K. It means that the type of magnetic order has no significant influence on the interdimer magnetic interaction. These findings are in qualitative agreement with the previous ambient-pressure study~\cite{Pantograph} that showed a reduction of approximately 10\% in the $J$ value, but no detectable changes of the interdimer coupling $J^\prime$ for different types of magnetic order. However, the values of the exchange couplings reported in the present paper are in much better agreement with the experiment.

\subsection{Pressure evolution}
In this section, we discuss the results of the structural optimization of SrCu$_2$(BO$_3$)$_2$ as a function of external pressure and magnetic order. All structural parameters reduce as pressure increases, which leads to a modification of the exchange couplings as will be discussed below.  The optimized structural parameters under pressure are listed in Table~\ref{tab:Structure} for the Dimer and N\'eel configurations, which are the relevant configurations in zero external field at lower and higher pressures, respectively. Since the spins on the dimers are parallel in both the ferromagnetic and N\'eel configurations, the structural parameters and ensuing magnetic interactions have similar values therein.    

\begin{table}[!h]
\caption [Bset]{Structural parameters of SrCu$_2$(BO$_3$)$_2$  calculated with N\'eel and Dimer magnetic orderings for different values of the external pressure. Experimental data  measured at ambient pressure and at low temperatures (T = 1.6 K~\cite{Zayed_PhD_diss} and T = 2 K~\cite{Vecchini}), and  under pressure P = 3.7 GPa at T = 300 K~\cite{Zayed_PhD_diss}   are given for comparison. 
 }
\begin{ruledtabular}
\setlength{\extrarowheight}{1pt}
\begin {tabular}{c|cccc}
   &  \multicolumn{2}{c}{$d$ (\AA)} &   \multicolumn{2}{c}{$\gamma$ (deg)} \\ 
 \hline
 P (GPa) & Dimer  & N\'eel  & Dimer & N\'eel   \\
 \hline
    0 ~\cite{Vecchini} &  2.931  & -   & 98.478  & -  \\ 
    0~\cite{Zayed_PhD_diss} & 2.922 & - &  97.851 & -\\
 
    0   &   2.897   &   2.891     &   97.470  &   97.186   \\
    1   &   2.889    &  2.883      &   97.346   &   97.068       \\
    2  &    2.882   &  2.875        &   97.257   &   96.983     \\
    3  &    2.875    &  2.868      &    97.194 &   96.926     \\

   3.7~\cite{Zayed_PhD_diss} &  - &  2.881    &  - &  94.520   \\  
    4 &   2.868   & 2.861   &   97.150   &   96.871     \\
 
\end {tabular}
\end{ruledtabular}
\label{tab:Structure}
\end {table}

\begin{figure}[!h]
\includegraphics[width=0.49\textwidth]{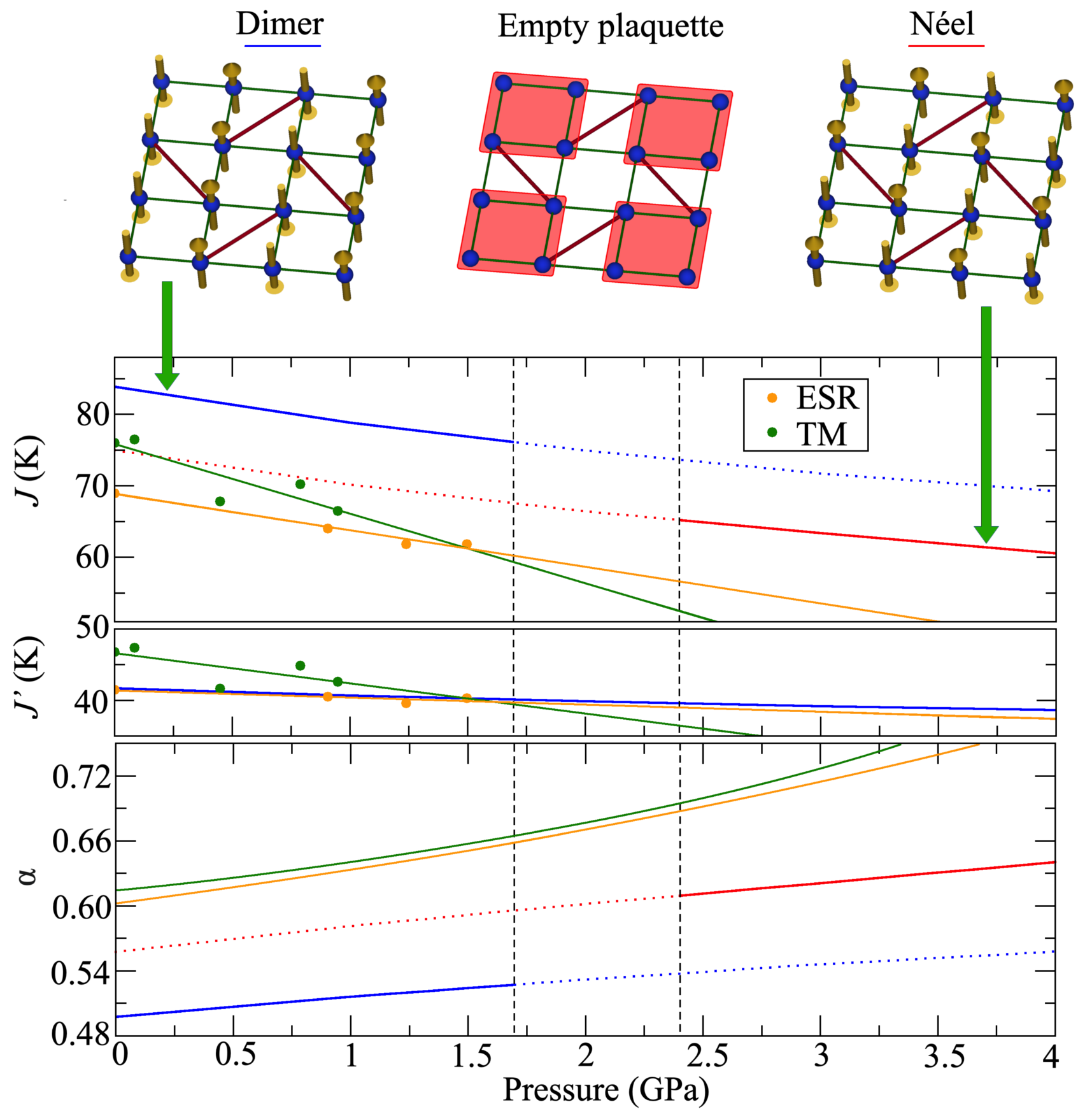}
\caption{ Pressure evolution of the intradimer $J$ and interdimer $J^\prime$ isotropic exchange interactions and their ratio $\alpha = J^\prime/J$. The interactions were estimated for the structures obtained with N\'eel and Dimer magnetic configurations.  The phase boundaries are taken from  Ref.~\onlinecite{Sandvik}. For comparison, we provide the available experimental estimates obtained from  thermodynamic measurements (TM)~\cite{Zayed} (green dots) and electron spin resonance (ESR)~\cite{Sakurai} (orange dots) data. Green and orange lines represent interpolation of the experimental data.}
\label{fig:Exchange_ALL}
\end{figure}

The exchange interactions as a function of pressure, $J(P)$ presented in Fig.~\ref{fig:Exchange_ALL}, have the same slope in the case of Dimer and  N\'eel orderings. By extrapolating the data to higher and lower pressures, respectively, we detect a finite shift between the corresponding $J$ values, which stems from the aforementioned spin-lattice coupling. One possible scenario is that $J(P)$ changes abruptly in the intermediate region. It is supported by the recent theoretical results reported in Ref.~\onlinecite{Nakano2018}. More specifically, the spin-spin correlation function $\braket{S^{z}_i S^{z}_j}$ for the shortest-distance copper pair coupled by the $J$ exchange interaction has a discontinuity at $\alpha_{c1} = 0.675$, the transition point to the intermediate phase~\cite{Nakano2018}. Due to the interplay between spin and lattice degrees of freedom, such a discontinuity of the  $\braket{S^{z}_i S^{z}_j}$ function can be expected to modify the equilibrium intradimer distance $d$ and the Cu -- O -- Cu angle $\gamma$, which will result in an abrupt variation of $J(P)$. The experimental estimates of the exchange interactions taken from Ref.~\onlinecite{Zayed, Sakurai} do not cover the intermediate phase, thus this is just a prediction at that stage.

At lower pressures, in the dimer phase, our ab-initio estimates of the exchange interactions are in better agreement with recent ESR results~\cite{Sakurai} than with those extracted from thermodynamic measurements~\cite{Zayed}. The slope of the pressure dependence is well reproduced, whereas deviations in the absolute values can be ascribed to a systematic error within DFT. This error is mostly affecting $J$ and, consequently, has impact on our computational estimate of $\alpha$ that comes out lower than 0.675 required for the breakdown of the singlet phase of the Shastry-Sutherland model~\cite{Miyahara}. 

\begin{figure}[!h]
\includegraphics[width=0.49\textwidth]{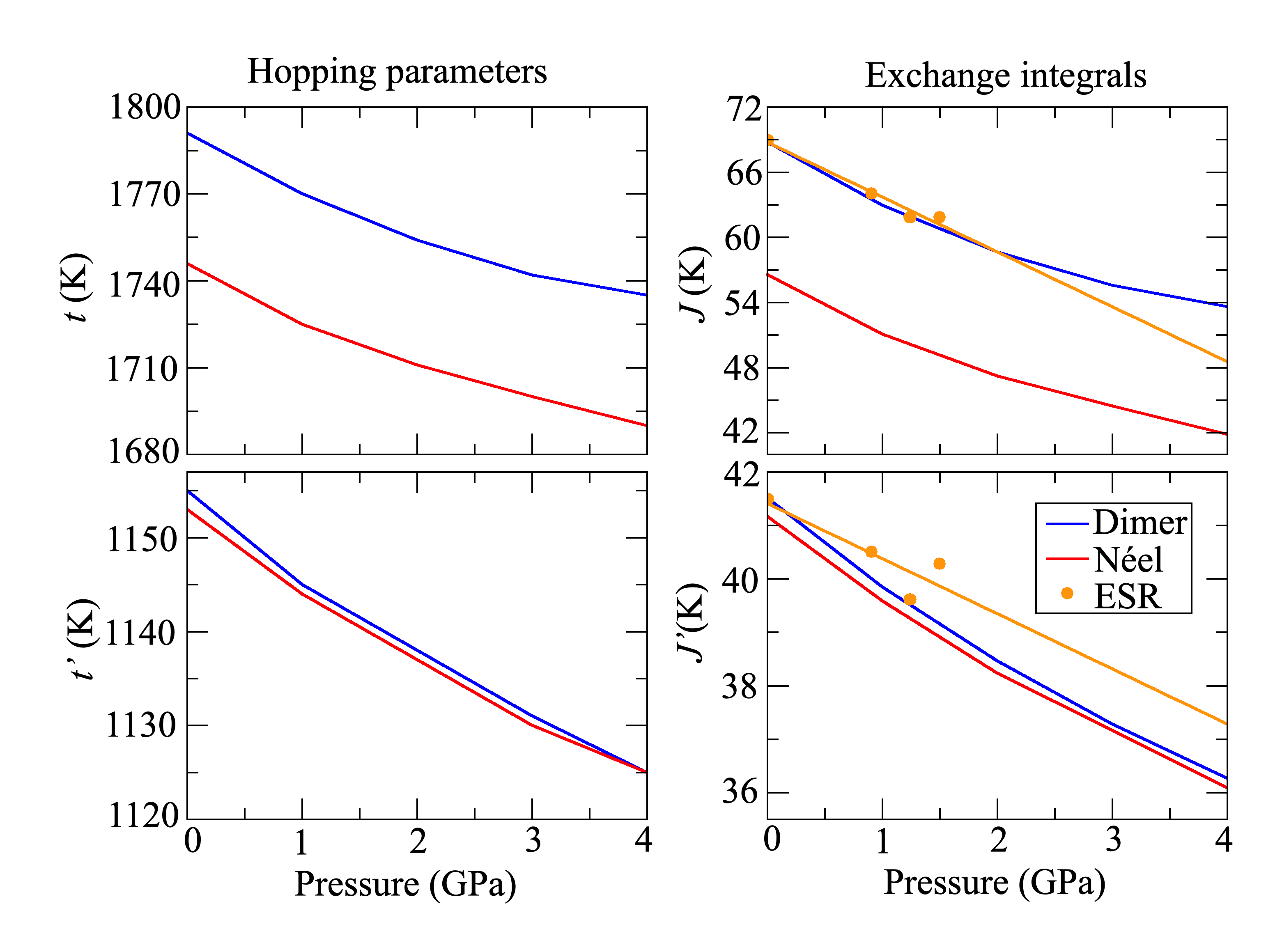}
\caption{ (Left) Pressure evolution of the intradimer $t$ and interdimer $t^\prime$ hopping parameters calculated in the basis of Wannier functions. (Right)   Exchange couplings,  obtained within Eq.~\ref{eq:Coupling}. The kinetic exchange interactions were calculated using hopping parameters, while the direct exchange couplings have been adjusted to fit the experimental  ESR~\cite{Sakurai} data (orange dots and line) at ambient pressure. Unlike the results of Fig.~\ref{fig:Exchange_ALL}, this calculation is thus not fully ab-initio, but it demonstrates the kinetic origin of the pressure dependence of the coupling constants.} 
\label{fig:Hopping}
\end{figure}

The intradimer coupling $J$ is a short-range coupling consisting of a ferromagnetic direct exchange and antiferromagnetic superexchange, see Eq.~\eqref{eq:Coupling}. Microscopically, the ferromagnetic part $J^F$ arises from Hund's coupling on the ligand site~\cite{mazurenko_wannier} and from the direct exchange $K_{pd}$ between copper $d$ and oxygen $p$ states~\cite{Matsuda2019, Stechel1988, Mizuno1998}. The former depends on the contribution of ligand orbitals to the Cu $d_{x^2-y^2}$ Wannier function~\cite{mazurenko_wannier}, which is nearly independent of pressure, as can be seen from the spread of the Wannier functions $W_{\mathbf{0}} (\mathbf{r})$,
\begin{eqnarray}
\Omega = [\braket{W_{\mathbf{0}} (\mathbf{r})| r^2 | W_{\mathbf{0}} (\mathbf{r})}  -  | \braket{W_{\mathbf{0}} (\mathbf{r})| \mathbf{r}| W_{\mathbf{0}} (\mathbf{r})}|^2 ].
\label{eq:Spread}
\end{eqnarray}
Calculations reveal that $\Omega$ (Table~\ref{tab:Spread}) decreases insignificantly within the pressure range of interest here. Regarding the $K_{pd}$ contribution, it scales with the $p-d$ hopping $t_{pd}$, which in turn depends on the Cu--O distance that shortens by 0.7\% at 4\,GPa. This will lead to a 1.4\% increase in $t_{pd}$ and 2.8\% increase in $J^F$, which is much smaller than the decrease in the antiferromagnetic part $J^{kin}$ due to the reduced hopping $t$. 
Therefore, to a first approximation pressure-induced changes in $J$ can be ascribed to the change in the hopping parameter and, consequently, in the Cu--O--Cu angle $\gamma$ according to $t \sim  \cos \gamma$ (Fig.~\ref{fig:Hopping})~\cite{mazurenko_wannier}.  The $\gamma$ value is systematically reduced under pressure, which decreases $J$, increases $\alpha$, and drives SrCu$_2$(BO$_3$)$_2$ through its pressure-induced magnetic transitions. DFT captures this trend very well, as can be seen from the perfect match between the slope of $J(P)$ obtained in our work and measured experimentally by ESR.  

\begin{table}[!h]
\caption [Bset]{Spreads $\Omega$ of Wannier functions of $x^2-y^2$ symmetry, calculated at different values of the external pressure and with different magnetic ordering (in \AA$^2$).}
\begin{ruledtabular}
\begin {tabular}{c|ccccc}
   &  0 GPa & 1 GPa & 2 GPa & 3 GPa & 4 GPa \\
   \hline
   N\'eel & 2.648 &  2.640	&  2.631  &  2.621  &   2.611 \\
   Dimer &   2.695 &  2.687  &  2.679  &  2.668 &  2.658  \\
\end {tabular}
\end{ruledtabular}
\label{tab:Spread}
\end {table}   

\begin{figure}[!h]
\includegraphics[width=0.49\textwidth]{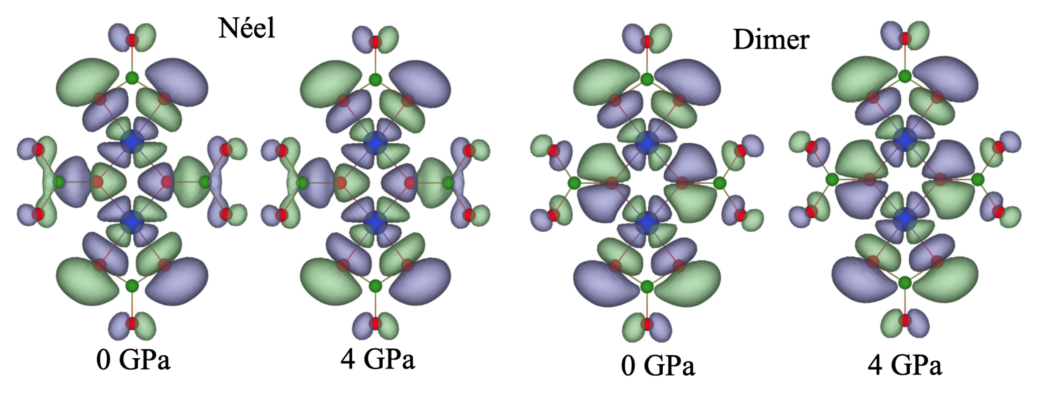}
\caption{Wannier functions centered at $x^2-y^2$ orbitals of nearest neighbour copper atoms. These functions were calculated with the crystal structures optimized at different pressures and magnetic configurations.}
\label{fig:WF}
\end{figure}

It is also useful to give numerical estimations of the different contributions to the total exchange interaction and compare their energy scales. To this end, we use the effective value of the Coulomb interaction $\widetilde{U}_{ij} = 4.5$ eV. At ambient pressure, the kinetic contribution to $J_{ij}$ in the case of Dimer order can be estimated as 246 K and 102 K for the intradimer and interdimer couplings, respectively.   Using these kinetic exchange interactions within Eq.~\eqref{eq:Coupling}, one can deduce the values of the direct exchange couplings, $J^{F}$ =  88.5 K and $J^{\prime F}$ = 30.5 K in order to fit available experimental data~\cite{Sakurai} obtained at ambient pressure.  Corresponding fitted curves reasonably describe experimental data under pressure, justifying  the main role of  kinetic exchange contribution in evolution of exchange couplings under pressure.

\subsection{\label{sec:Pantograph} Pantograph mode}
Previous experimental studies suggested a strong spin-lattice coupling in SrCu$_2$(BO$_3$)$_2$~\cite{Kodama2002}. This motivates the modeling of phonon spectra using different magnetic configurations, as well as crystal structures optimized under pressure.
\begin{figure}[!h]
\centering
\includegraphics[width=0.49\textwidth]{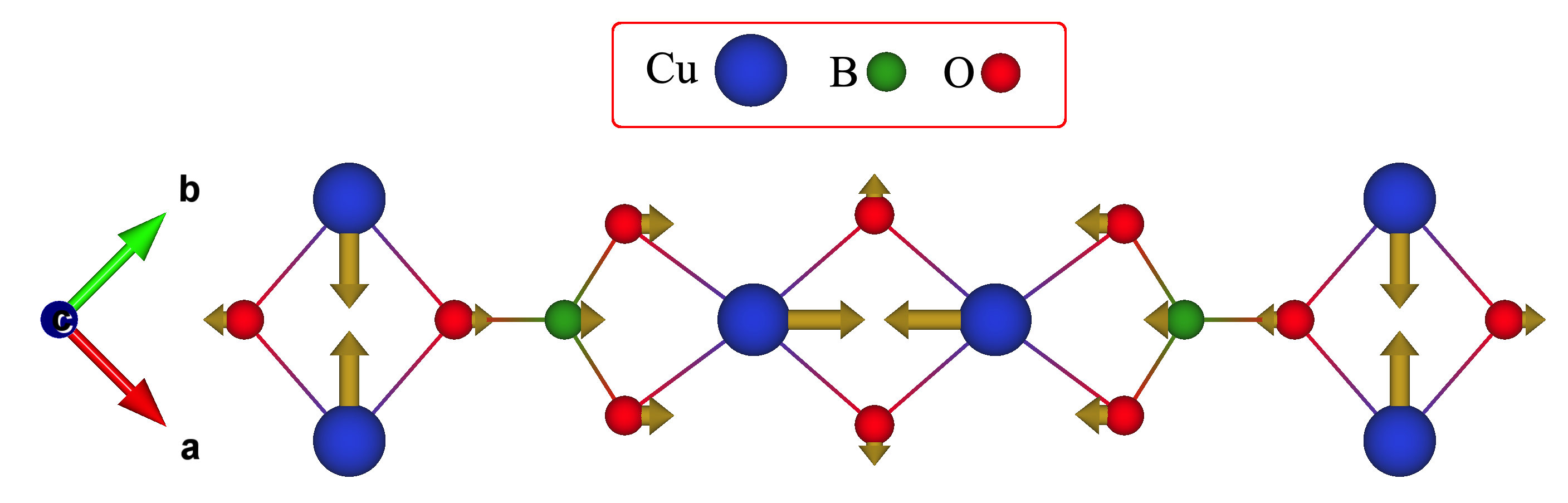}
\caption{ Pantograph mode atomic displacements according to DFT+$U$ simulations. The sizes of the arrows are proportional to the displacement vectors obtained from the calculations. See Fig.~\ref{fig:Crystal} for more information about the crystal structure of SrCu$_2$(BO$_3$)$_2$. }
\label{fig:Pantograph}
\end{figure}

Our focus is on the so-called pantograph mode, schematically represented in Fig.~\ref{fig:Pantograph}. It directly modulates the dimer bond length $d$ and the Cu -- O -- Cu angle $\gamma$, thus modifying the intradimer coupling $J$~\cite{Pantograph}. Recently, using Raman scattering it was shown that this mode exhibits a very strong pressure and temperature dependence~\cite{Zheludev2019}. The relative frequency shift of the pantograph mode between 40 K and 2.6 K shows an abrupt change at P$_1 \sim $ 1.5 GPa and gradually increases, switching sign at about P$_2 \sim $ 2.2 GPa. Since the pantograph mode is related to the intradimer geometry, the observed pressure dependence can be related to the dimer bond energy $J \braket{\hat{\mathbf{S}}_1\cdot  \hat{\mathbf{S}}_2}$. Thus, the estimated critical pressure values can be expected to reflect the transitions from dimer singlet to the intermediate state at $P_1$ and from intermediate phase to N\'eel at $P_2$ reported in recent high-pressure thermodynamic measurements~\cite{Sandvik}.

 \begin{figure}[!h]
\centering
\includegraphics[width=0.49\textwidth]{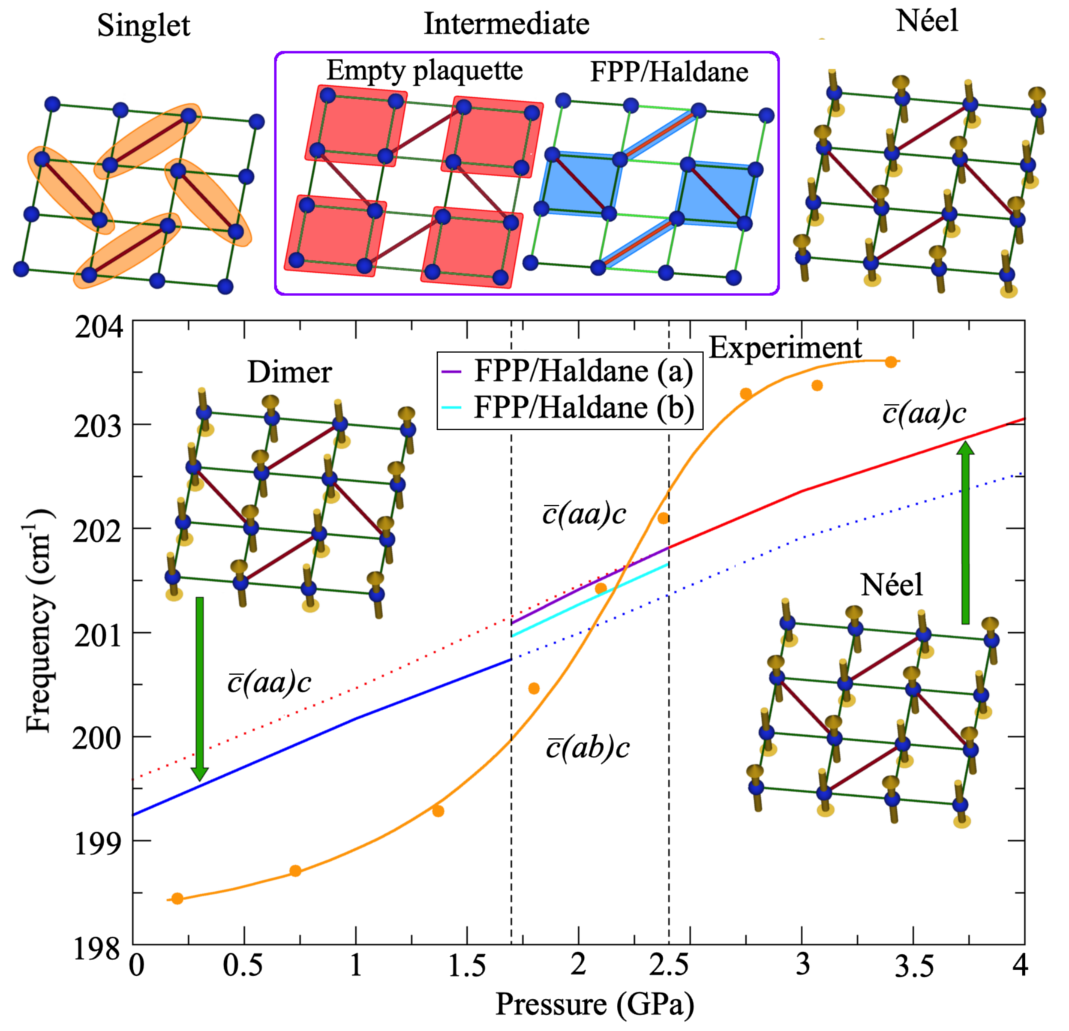}
\caption{ Pressure evolution of the pantograph mode frequency: comparison of the phonon mode calculated for Dimer and N\'eel types of  magnetic order  within DFT+$U$, with the experimental data measured at 2.6 K~\cite{Zheludev2019}.  The phase boundaries are taken from  Ref.~\onlinecite{Sandvik}. The frequencies within the FPP/Haldane phases are given for comparison (see Sec.~\ref{sec:Haldane}). }
\label{fig:Pantograph_evolution}
\end{figure} 

The calculated  phonon frequencies with pantograph symmetry for Dimer and N\'eel magnetic configurations and their evolution under pressure are represented in Fig.~\ref{fig:Pantograph_evolution}.  Similarly to the intradimer coupling $J(P)$, the frequencies for both configurations are characterized by the same slope with a finite shift between them. For the Dimer order, which simulates the magnetic ground state at low pressures, we observe smaller values of the frequencies in comparison with the high-pressure N\'eel state.  This difference is significant and related to the different $d$ and $\gamma$ values in the two structures. The shortening of $d$ leads to the hardening of the pantograph mode. However, the estimated shift between the frequencies in the Dimer and N\'eel phases is significantly smaller than that measured in experiment, which means that DFT may underestimate the spin-lattice coupling or does not fully reproduce the nature of the magnetic state above 2\,GPa. In addition, the change in the pantograph mode frequency at $\sim$ 2 GPa can also have a structural contribution on top of the magnetic one, as further discussed in Sec.~\ref{sec:Haldane}.

\section{\label{sec:Haldane} Intermediate magnetic phase }

\subsection{FPP/Haldane-type candidate }
The realization of the intermediate empty plaquette phase (Fig.~\ref{fig:SS_model}) implies that the system preserves its tetragonal symmetry under pressure. However, this is not the only possibility.  The magnetic lattice of  SrCu$_2$(BO$_3$)$_2$ is highly frustrated, and the frustration degree is controlled by the ratio $\alpha = J^\prime /J$: The degree of frustration is higher in the intermediate phase between the unfrustrated limits of decoupled singlets and square-lattice antiferromagnet.  In order to release such a frustration, the system may experience  some distortions of the tetragonal crystal structure under pressure. This can be studied with DFT to complete existing experimental data.  

\begin{figure}[!h]
\includegraphics[width=0.49\textwidth]{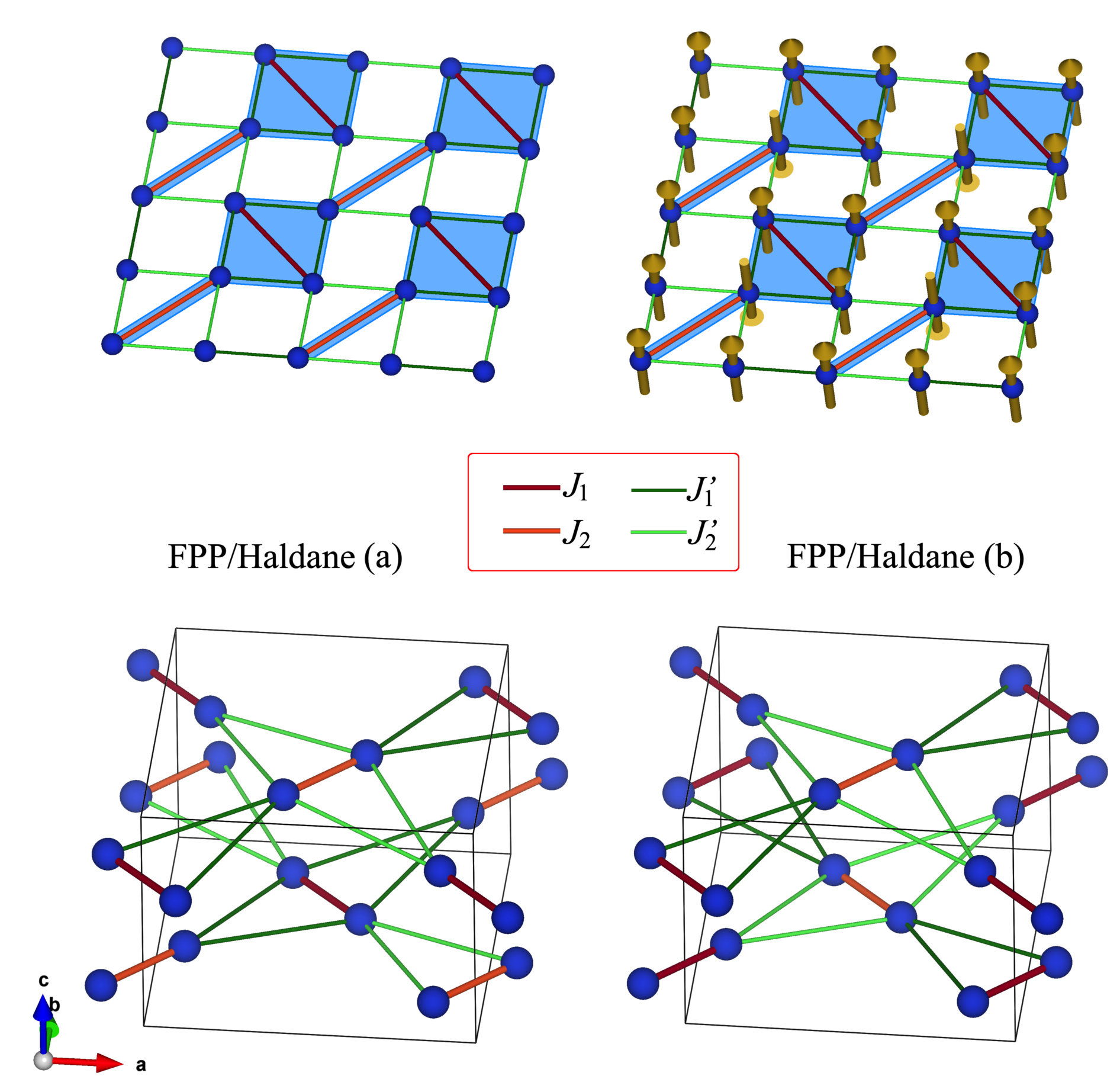}
\caption{(Top) FPP/Haldane magnetic state  and its representation with the magnetic configuration used in this work.
(Bottom) Crystal structure of SrCu$_2$(BO$_3$)$_2$, where in FPP/Haldane (a)  AFM ordered dimers are oriented in the same direction for both structural layers, and in FPP/Haldane (b) this direction alternates from one layer to the next. }
\label{fig:Haldane}
\end{figure}

One of the possible distortion scenario has been proposed by Boos $et$  $al.$~\cite{Boos}. The authors have studied an extended Shastry-Sutherland model with nonequivalent intradimer $J_1$, $J_2$ and interdimer $J_1^\prime$, $J_2^\prime$ couplings. It was found that if $J_1$ and $J_2$ are sufficiently different, it is possible to stabilize the so-called full plaquette (FPP) phase. By contrast to the empty plaquette phase (EPP), which is known to be realized in the intermediate phase of the Shastry-Sutherland model (Fig.~\ref{fig:SS_model}), the FPP is characterized by stronger correlations around plaquettes with diagonal couplings, effectively forming chains in the two-dimensional magnetic lattice  (Fig.~\ref{fig:Haldane}), and it is adiabatically connected to the Haldane magnetic phase suggested in Ref.~\onlinecite{Moliner}. The calculated structure factor agrees with neutron scattering experimental data under pressure~\cite{Zayed}. Moreover, numerical studies demonstrate that the spin gap $\Delta$ increases as $\alpha$ increases in the case of the EPP, while the corresponding FPP/Haldane magnetic phase demonstrates a reduction of $\Delta$~\citep{ Boos}, in agreement with recent specific heat measurements under pressure~\cite{Sandvik}.

Thus the next important step of our investigation is to choose the best representation of the FPP/Haldane magnetic phase of SrCu$_2$(BO$_3$)$_2$ in terms of a mean-field magnetic configuration to be studied within  DFT. To that end, we will consider a magnetic order that is characterized by ferromagnetic dimers in one crystallographic direction and antiferromagnetic dimers in the orthogonal in-plane direction (Fig.~\ref{fig:Haldane}). In this way, FM dimers will behave as effective magnetic sites with spin $S$ = 1 interacting through AFM bonds. 
As shown in Sec.~\ref{sec:Tetragonal}, FM and AFM ordered dimers have different Cu -- Cu distances $d$ and Cu -- O -- Cu angles $\gamma$, which we label as $d^{AFM/FM}$ and  $\gamma^{AFM/FM}$, respectively. Thus, it is expected that such a magnetic order will break the initial tetragonal symmetry.

In the case of the three-dimensional magnetic structure, there are two possibilities for the magnetic order within the unit cell: AFM ordered dimers are oriented in the same direction or alternate from one layer to the next. Depending on this, the crystal structure will be modified accordingly.  

\subsection{Structural properties}

Let us consider the situation where all AFM ordered dimers are parallel to the  $\mathbf{a} + \mathbf{b}$ direction in both layers of the unit cell. In  Fig.~\ref{fig:Haldane}, such an order is labeled as FPP/Haldane (a).

\begin{table}[!h]
\caption [Bset]{Optimized structural parameters of SrCu$_2$(BO$_3$)$_2$ with FPP/Haldane (a) magnetic order (Fig.~\ref{fig:Haldane}). }
\begin{ruledtabular}
\setlength{\extrarowheight}{1pt}
\begin {tabular}{c|cccc}
   P (GPa) & $d^{AFM}_1$ (\AA) &  $d^{FM}_2$ (\AA)  & $\gamma^{AFM}_1$ (deg) &  $\gamma^{FM}_2$ (deg)  \\
 \hline
 0  &   2.896 &  2.891   & 97.439 & 97.215  \\  
  1  &   2.889 & 2.883  & 97.317 & 97.096  \\
 2  &   2.882 &  2.876  & 97.233 & 97.008  \\
 3  &   2.875 &  2.869  & 97.172 & 96.953  \\
 4  &   2.868 &  2.862  &  97.120 & 96.904   \\
\end {tabular}
\end{ruledtabular}
\label{tab:Haldane_a}
\end {table}

The structural parameters optimized at different pressure values are summarized in Table~\ref{tab:Haldane_a}. The atomic positions at ambient pressure are given in Appendix~\ref{app:Structure}.  Importantly, this magnetic order leads to the stabilization of the structure with two nonequivalent dimer distances, where $\Delta d  =  d^{AFM} - d^{FM} \sim$ 0.006 \AA ,\ and Cu -- O -- Cu angle $\Delta \gamma = \gamma^{AFM} - \gamma^{FM}  \sim$ 0.22$^\circ$ in the entire pressure range, similarly to the difference between pure FM and Dimer ordering (see Table.~\ref{tab:Structure}). Consequently, the system has a distortion along the  $\mathbf{a} + \mathbf{b}$ direction, since AFM ordered dimers with the longer intradimer $d$ distance are oriented in the same direction in both layers. Thus, the lattice vectors $\mathbf{a}$ and $\mathbf{b}$ are not orthogonal to each other any more. The symmetry analysis shows that such a crystal structure belongs to the $Cm$ space group (No. 8).

\begin{table}[!h]
\caption [Bset]{Structural parameters of SrCu$_2$(BO$_3$)$_2$ optimized at different values of external pressure. These results were obtained with the FPP/Haldane (b) magnetic order (Fig.~\ref{fig:Haldane}). }
\begin{ruledtabular}
\setlength{\extrarowheight}{1pt}
\begin {tabular}{c|cccc}
   P (GPa) & $d^{AFM}_1$ (\AA) &  $d^{FM}_2$ (\AA)  & $\gamma^{AFM}_1$ (deg) &  $\gamma^{FM}_2$ (deg)  \\
 \hline
 0  &   2.895   &  2.892   & 97.398 & 97.255  \\
 1  &   2.888 &   2.885  & 97.276  & 97.134   \\
  2  &   2.881  &   2.878  & 97.190  & 97.049  \\
   3  &   2.873  &  2.870  & 97.132  & 96.993  \\
   4  &   2.867  &  2.864  & 97.084  & 96.945  \\
\end {tabular}
\end{ruledtabular}
\label{tab:Haldane_b}
\end {table}

The magnetic structure considered above is only one of the possible representations consistent with a many-body FPP/Haldane phase in each layer. Another possibility is that the AFM dimers alternate in the crystal structure. Such a configuration is denoted as FPP/Haldane (b) in Fig.~\ref{fig:Haldane}. In this case, the $ab$ plane distortions are eliminated (Table~\ref{tab:Haldane_b}) because the AFM dimers with long $d$ distance are orthogonal between neighbouring layers, leading to a compensation of the elongation of the unit cell in one specific direction. This alternation results in twice smaller differences $\Delta d \sim$  0.003 \AA ,\ and $\Delta \gamma \sim$ 0.14$^\circ$.  In this case, the system is characterized by the $P2$ space group (No. 3), i.e. the symmetry is lower than in the FPP/Haldane (a) phase. 

\subsection{Exchange interactions}

\begin{figure}[!h]
\includegraphics[width=0.49\textwidth]{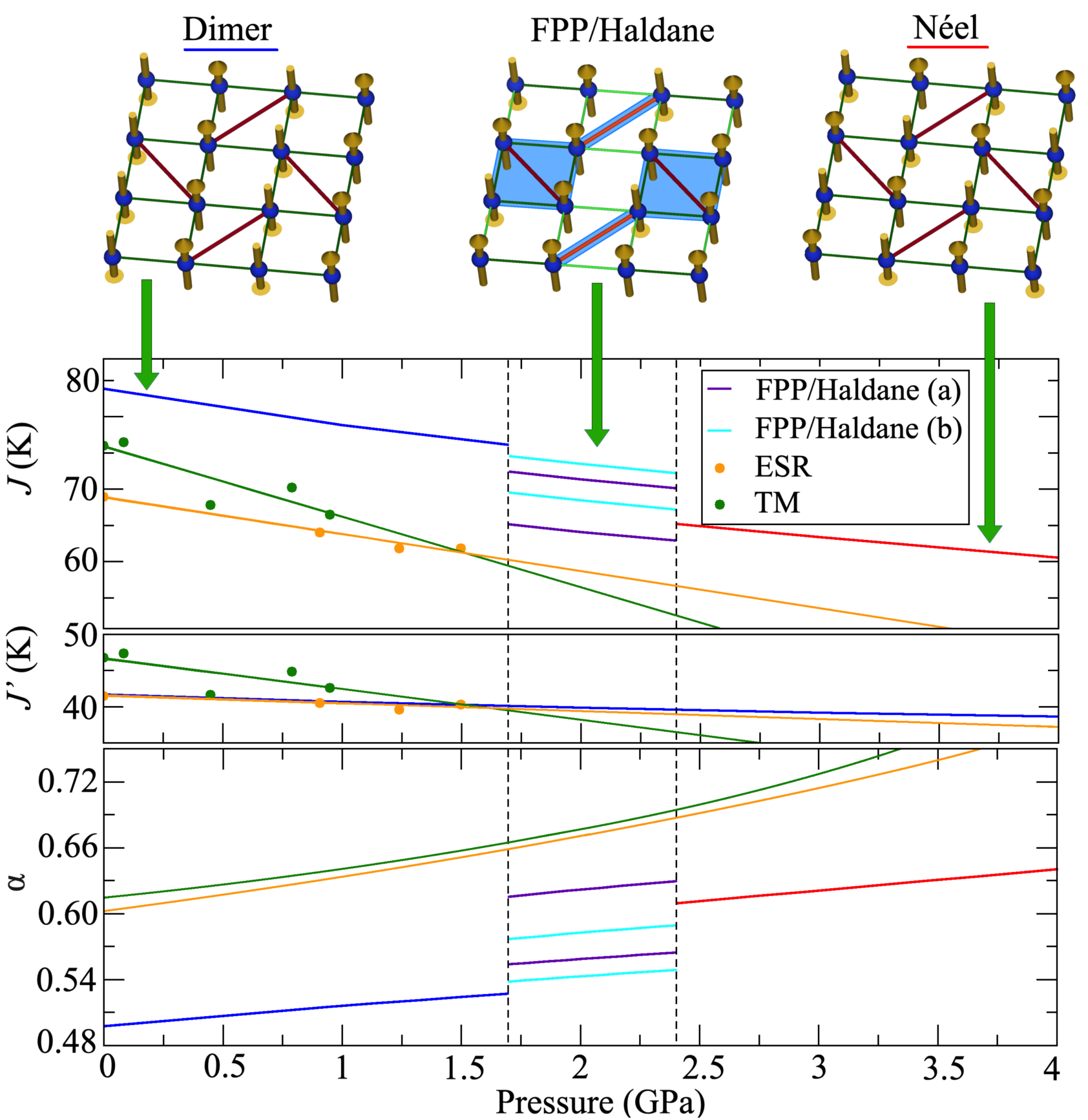}
\caption{Pressure evolution of the inequivalent intradimer interactions $J_1$ (FM  ordered pair) and $J_2$ (AFM  ordered pair), and of the interdimer  isotropic exchange interactions $J_{1-2}^\prime$ calculated for the two possible magnetic configurations for the FPP/Haldane magnetic phase (Fig.~\ref{fig:Haldane}). The phase boundaries are taken from Ref.~\onlinecite{Sandvik}. For comparison, we provide the available experimental estimates obtained from thermodynamic measurements (TM)~\cite{Zayed} and electron spin resonance (ESR)~\cite{Sakurai} data.}
\label{fig:Exchange_Haldane}
\end{figure}

The calculated values of the exchange interactions are given in Fig.~\ref{fig:Exchange_Haldane}.  It is interesting to note that the magnetic configurations representing the FPP/Haldane phases lead to rather different values of $J_1$ and $J_2$. In the FPP/Haldane (a) phase, the smallest value of $J_1$ is outside the window obtained by extrapolating the values of $J(P)$ in the Dimer and in the N\'eel phases, while in the FPP/Haldane (b) phase, both $J_1$ and $J_2$ are inside that window.

The magnetic configurations representing the FPP/Haldane phases should also affect the interdimer couplings $J^\prime$.   Indeed, in the $J_2 > J_1$ case, the relation $J^\prime_2 < J^\prime_1$ should hold because the increase in $J$ through the increasing Cu -- O -- Cu angle $\gamma$ will lead to a decrease  in the Cu -- O -- O angles of the Cu--O--O--Cu superexchange pathway and may eventually reduce $J^\prime$. However, despite the significant difference in the intradimer couplings ($J_1/J_2 \sim $ 0.90 in  FPP/Haldane (a)   and $J_1/J_2 \sim $ 0.93 in  FPP/Haldane (b) phases (Fig.~\ref{fig:Haldane})), we find $J^{\prime}_1/J^{\prime}_2 \sim 1$ in both cases, meaning that the interdimer couplings only weakly depend on the Cu -- O -- O -- Cu angle. On the other hand, even equal values of these couplings ($J_2'/J_1'=1$) should be sufficient for the stabilization of the FPP/Haldane phase~\cite{Boos} if $J_1/J_2\simeq 0.9$, which is the case in our calculations.

\subsection{Experimental signatures}
The structures of the FPP/Haldane phases feature lower symmetry than pristine SrCu$_2$(BO$_3)_2$. By calculating x-ray structure factors for the optimized crystal structures, we conclude that the FPP/Haldane (a) phase should be characterized by a weak peak splitting, with the largest peak separation of $\Delta d \sim 3\times 10^{-3}$\AA \, for the $110$ reflection of tetragonal SrCu$_2$(BO$_3)_2$. The FPP/Haldane (b) phase features a larger unit cell due to the absence of $C$-centering and will thus show additional  reflections that do not occur in the tetragonal SrCu$_2$(BO$_3)_2$. All these effects are rather weak, though, which may explain why no signature of structural distortions was observed in previous diffraction experiments under pressure. Nuclear Quadrupolar Resonance is also sensitive to structural distortions~\cite{Waki2007, Takigawa2010}, but so far this method did not reveal any distortion in the intermediate phase.

 Alternatively, structural distortions can be detected by Raman scattering under pressure.  In particular, the symmetry analysis of the undistorted tetragonal structure with the space group $I \overline{4}  2m$ gives the following representation of the optical modes: 
\begin{equation}
\Gamma_{optic} = 9A_1 + 7A_2 +6B_1 + 9B_2 +16E.
\end{equation}
All the modes except $A_2$ are Raman active, while only $B_2$ and $E$ are infra-red (IR) active. According to the selection rules for Raman scattering, only modes with specific symmetry can be observed for a given polarization. Structural distortions may lead to a violation of such selection rules,  implying that modes become observable in several polarizations, which would not be possible in the tetragonal phase. 

 In particular, the pantograph mode analyzed in Sec.~\ref{sec:Pantograph} has $A_1$ symmetry in the tetragonal phase. It can be observed in the $\overline{c}(aa)c$ polarization~\cite{Zheludev2019}, but remains silent in the $\overline{c}(ab)c$ polarization. The magnetic configurations (a) and (b) of FPP/Haldane type modify the symmetry of this mode to $A^{'}$ and $A$, respectively, activating it for both $\overline{c}(aa)c$ and $\overline{c}(ab)c$ polarizations. Thus, this mode will not allow to determine the type of distortion (FPP/Haldane (a) or (b)), but it could give direct evidence that the distortion occurs if such a selection rule violation takes place (Fig.~\ref{fig:Pantograph_evolution}). Thus, it would be  highly desirable to look for violations of the selection rules in Raman scattering experiments under pressure.

\section{\label{sec:conclusions}Conclusions}
In this work, we studied the structural and magnetic properties of SrCu$_2$(BO$_3$)$_2$ under pressure by using density-functional-theory methods. Depending on the pressure value, different mean-field magnetic configurations approximating quantum magnetic states were considered:  
i) Dimer for the singlet state at low pressures, in which nearest-neighbor copper pairs are exact singlets; ii) ferromagnetic for the fully polarized state that would be realized in a high magnetic fields; iii) N\'eel for the high-pressure phase in which the intradimer correlations are ferromagnetic (as the diagonal correlations in the N\'eel state on the square lattice); and iv) magnetic configurations of FPP/Haldane type in the intermediate pressure phase in which half of the nearest-neighbor dimers have antiparallel spins and the remaining half have parallel spins. In all cases, we observed that the choice of a particular magnetic order modifies the crystal structure.  Spin-lattice coupling effects at ambient pressure -- the stretching of the dimers upon cooling and their contraction in the applied magnetic field -- have been reproduced, suggesting that mean-field magnetic configurations capture the spin-spin correlations and the associated bond energies of the quantum spin state.

Hydrostatic pressure increases the ratio $\alpha = J^\prime / J $ and triggers changes in the magnetic ground state toward the magnetically ordered phase through an intermediate phase. We explored possible structural distortions within this phase and detected  two nonequivalent dimers with different couplings $J_1$ and $J_2$, with a $J_1 / J_2$ ratio lying in the range of the FPP/Haldane magnetic state in the phase diagram of the extended Shastry-Sutherland model~\citep{Boos}. Finally, on the basis of the symmetry analysis of the calculated phonon modes, we discuss possible ways to detect this intermediate phase in Raman scattering experiments.

\section{\label{Acknowledgements}Acknowledgements}
D.I.B. acknowledges the Russian Federation Presidential scholarship for providing travel support to visit EPFL. The work of V.V.M. is supported by the Russian Science Foundation, Grant No. 18-12-00185. A.A.T. acknowledges financial support by the Federal Ministry for Education and Research through the Sofja Kovalevskaya Award of Alexander von Humboldt Foundation. We acknowledge A. Zheludev and S. Bettler (ETH Z\"urih) for fruitful discussions and sharing Raman scattering data under pressure prior to publication. The calculations have been performed using the facilities of the Scientific IT and Application Support Center of EPFL. F.M. acknowledges the support of the Swiss National Science Foundation.

\appendix
\section{ \label{app:Structure}Atomic positions for FPP/Haldane magnetic states}

Here we provide the atomic positions, optimized  within DFT+$U$ method ($U$ = 11.2  eV, $J_H$ = 1 eV) without external pressure. The magnetic order during the optimization was fixed as FPP/Haldane (a) and (b) (Fig.~\ref{fig:Haldane}).

\begin{table}[!h]
\caption [Bset]{Structural parameters of  SrCu$_2$(BO$_3$)$_2$, optimized  at ambient pressure with FPP/Haldane (a) magnetic order (Fig.~\ref{fig:Haldane}).}
\begin{ruledtabular}
\begin {tabular}{ccccc}
\multicolumn{5}{c}{$a$ = 7.2248 \AA, $b$ = 12.7235 \AA,  $c$ = 6.8363 \AA} \\
\multicolumn{5}{c}{$\alpha$ = $\gamma$ = 90$^\circ$, $\beta$ = 118.236$^\circ$} \\
\multicolumn{5}{c}{Space group  is  $Cm$ (No. 8)} \\
\hline
   B1         & $2a$   &    -0.5874   &     0    &    -0.5338    \\
   B2         & $2a $  &   -0.4125   &     0    &    0.0536     \\
   B3         & $4b$  &     0.0000   &     0.2063    &    0.7401      \\
   Cu1       & $2a$   &   -0.2275   &     0    &     0.5973     \\
   Cu2       & $2a$   &   -0.7725   &     0    &   -0.1753      \\
   Cu3       & $4b$   &     0.0000   &     0.3864    &    0.7890      \\
   O1         & $2a$    &   -0.8002   &      0   &    -0.6010    \\
   O2         & $2a$   &   -0.1998   &      0   &     0.1991     \\
   O3         &  $4b$  &   0.0000   &      0.0998   &     0.7012     \\
   O4         &  $4b$  &   -0.4696  &      0.4088   &     0.0065    \\
   O5         &  $4b$  &   -0.5303   &      0.0912    &   -0.0238    \\
   O6         &  $4b$  &    -0.8177    &      0.2652    &   -0.1502    \\
   O7         &  $4b$  &    -0.1821    &      0.2348   &    0.1674      \\
   Sr1        &   $4b$  &    -0.5000    &      0.2500   &   -0.2500     \\
\end{tabular}
\end{ruledtabular}
\label{tab:Haldane_a2}
\end{table}

\begin{table}[!h]
\caption [Bset]{Structural parameters of  SrCu$_2$(BO$_3$)$_2$, optimized  at ambient pressure with FPP/Haldane (b) magnetic order (Fig.~\ref{fig:Haldane}).}
\begin{ruledtabular}
\begin {tabular}{ccccc}
\multicolumn{5}{c}{$a$ = $b$ =  8.9993 \AA, $c$ = 6.8362 \AA} \\
\multicolumn{5}{c}{$\alpha$ = $\beta$ =  $\gamma$ = 90$^\circ$} \\
\multicolumn{5}{c}{Space group  is  $P_2$ (No. 3)} \\
\hline
   B1      & $2e$        &     0.7063   &   0.2937   &    0.7600    \\
   B2     & $2e$        &     0.2937    &   0.7063   &    0.7599    \\
   B3     & $2e$        &     0.2063   &   0.7938    &    0.2598   \\
   B4     & $2e$        &     0.7938    &   0.2063   &     0.2599   \\
   Cu1   & $2e$        &     0.8862    &   0.1137     &     0.7111     \\
   Cu2   & $2e$        &     0.1137      &   0.8863   &     0.7111     \\
   Cu3   & $2e$        &     0.3864    &    0.6136    &     0.2110   \\
   Cu4   & $2e$        &     0.6136     &    0.3864    &    0.2109    \\
   O1     & $2e$         &     0.6000   &    0.4000     &    0.7993   \\
   O2     & $2e$        &     0.4001    &     0.5998    &    0.7990  \\
   O3    & $2e$         &     0.0998    &    -0.0998   &     0.2986  \\
   O4     & $2e$        &    -0.0998   &     0.0998    &    0.2988   \\
   O5     & $2e$        &     0.6740    &     0.1437     &     0.7414   \\
   O6     & $2e$        &     0.3260   &    0.8563       &     0.7414   \\
   O7      & $2e$       &     0.8563    &    0.6740      &    0.2587   \\
   O8      & $2e$       &      0.1437  &     0.3260      &     0.2586   \\
   O9      & $2e$       &     0.1741    &     0.6437      &     0.2415    \\
   O10    & $2e$       &     0.8260   &     0.3564      &     0.2413   \\
   O11     & $2e$       &     0.3563   &     0.1740       &    0.7587    \\
   O12     & $2e$       &     0.6436   &     0.8259     &    0.7585    \\
   Sr1      & $1a$       &     0   &     0.5000     &    0     \\
   Sr2     & $1c$        &    1/2    &     0.0000    &    0    \\
   Sr3     & $1d$        &     1/2   &     0.0000    &  1/2    \\
   Sr4     & $1b$       &     0   &     0.5000     &    1/2   \\
\end{tabular}
\end{ruledtabular}
\label{tab:Haldane_b2}
\end{table}


%

\end{document}